\documentclass[MSNbibl,nameyear,dvips]{arxstspdf}
\usepackage{dcolumn}
\usepackage{graphicx}
\usepackage{flushend}
\usepackage{stfloats}

\volume{28}
\issue{3}
\pubyear{2013}
\firstpage{360}
\lastpage{375}
\doi{10.1214/13-STS423} 

\makeatletter
\newcolumntype{d}[1]{D{.}{.}{#1}}

\newcommand{\RQ}{\mathrm{RQ}}

\newcommand{\PV}{\mathit{PV}}

\newtheorem{theorem}{Theorem}
\newtheorem{corollary}{Corollary}

\newproclaim{example}{Example}

\newcommand{\sX}{\mathsf{X}}

\makeatother

\begin{document}
\begin{frontmatter}

\title{Component-Wise Markov Chain Monte Carlo: Uniform and Geometric
Ergodicity under Mixing and Composition}
\runtitle{Component-Wise Markov Chain Monte Carlo}

\begin{aug}
\author[a]{\fnms{Alicia A.} \snm{Johnson}\ead[label=e1]{ajohns24@macalester.edu}},
\author[b]{\fnms{Galin L.} \snm{Jones}\ead[label=e2]{galin@stat.umn.edu}}
\and
\author[c]{\fnms{Ronald C.} \snm{Neath}\corref{}\ead[label=e3]{rneath@hunter.cuny.edu}}
\runauthor{A. A. Johnson, G. L. Jones and R. C. Neath}

\affiliation{Macalester College, University of Minnesota and City University of New York}

\address[a]{Alicia A. Johnson is Assistant Professor,
Department of Mathematics, Statistics,
and Computer Science, Macalester College,
Saint Paul, Minnesota 55105, USA \printead{e1}.}
\address[b]{Galin L. Jones is Associate Professor, School of Statistics,
University of Minnesota,
Minneapolis, Minnesota 55405, USA \printead{e2}.}
\address[c]{Ronald C. Neath is Assistant Professor, Department of Mathematics and
Statistics,
Hunter College, City University of New York,
New York, New York 10065, USA \printead{e3}.}

\end{aug}

%
\begin{abstract}
It is common practice in Markov chain Monte Carlo to update the
simulation one variable (or sub-block of variables) at a time,
rather than conduct a single full-dimensional update. When it is
possible to draw from each full-conditional distribution associated
with the target this is just a Gibbs sampler. Often at least one of
the Gibbs updates is replaced with a Metropolis--Hastings step,
yielding a Metropolis--Hastings-within-Gibbs algorithm. Strategies
for combining component-wise updates include composition, random
sequence and random scans. While these strategies can ease MCMC
implementation and produce superior empirical performance compared
to full-dimensional updates, the theoretical convergence properties
of the associated Markov chains have received limited attention. We
present conditions under which some component-wise Markov chains
converge to the stationary distribution at a geometric rate. We pay
particular attention to the connections between the convergence
rates of the various component-wise strategies. This is important
since it ensures the existence of tools that an MCMC practitioner
can use to be as confident in the simulation results as if they were
based on independent and identically distributed samples. We
illustrate our results in two examples including a hierarchical
linear mixed model and one involving maximum likelihood estimation
for mixed models.
\end{abstract}

%

\end{frontmatter}

\section{Introduction}
\label{secIntro}
Let $\varpi$ be a probability distribution having support $\mathsf{X}
\subseteq\mathbb{R}^{q}$, $q \ge1$. The fundamental Markov chain
Monte Carlo (MCMC) method for making draws from $\varpi$ is the
Metropolis--Hastings algorithm, described here. Let $X^{(k)} = x$
denote the current state, and suppose $\varpi$ has a density function
$\pi$. Let $p(\cdot, \cdot)$ denote the user-defined proposal
density. The updated state $X^{(k+1)}$ is obtained via the following:
\begin{enumerate}
\item Simulate $x^*$ from proposal density $p(x, \cdot)$.
\item
Calculate acceptance probability $\alpha(x,x^*)$, where
\[
\alpha(x,y) = \min \biggl\{ 1, \frac{\pi(y)}{\pi(x)} \frac{p(y,x)} {
p(x,y)} \biggr\}.
\]
\item Set
\[
X^{(k+1)} = \cases{ x^*, & with probability $\alpha\bigl(x,x^*\bigr)$,
\cr
x, & with probability $1 -\alpha\bigl(x,x^*\bigr)$.}
\]
\end{enumerate}
Thus, creating a Metropolis--Hastings sampler boils down to choosing a
proposal density $p(\cdot, \cdot)$. If\break $p(x, y) = p(y, x)$, this is a
\textit{Metropolis} algorithm. If, further, $p(x,y) = p(x - y) = p(y
- x)$ for all $x$ and~$y$, it is a \textit{Metropolis random
walk}. When the proposal $p(\cdot)$ does not depend on the current
state the chain is a \textit{Metropolis--Hastings independence
sampler}\break
(MHIS).

The selection of the proposal density can be challenging, particularly
in problems where $q$ is large or the support of $\varpi$ is
complicated. This has led to investigation of optimal scaling of
Metropolis algorithms and so-called adaptive algorithms which allow
the proposal kernel to change over the course of the simulation
(see, e.g., \cite{bedarose2008}; \cite{rose2011}). An
alternative to full-dimensional updates is a \textit{component-wise}
approach where we update one variable (or sub-block of variables) at a
time.

The choice between full-dimensional and compo\-nent-wise updates is
frequently unclear (see, e.g., \cite{robesahu1997}), although a
general guideline seems to be that updating as a single block may not
be advantageous if the components of $\varpi$ are only weakly
correlated. For example, \citet{nealrobe2006} considered target
distributions having independent and identically distributed components
and showed that for an idealized version of a component-wise
Metropolis algorithm it is optimal to update only one variable at a
time. On the other hand, these authors also showed that when using
Metropolis adjusted Langevin algorithms, block updating is more
efficient.

Whatever MCMC method is used, an important consideration is the rate
of convergence of the chain to its stationary distribution. Let
$\mathcal{B}$ be the Borel $\sigma$-algebra on $\mathsf{X}$ and let
$P^n(x, dy)$ denote the $n$-step Markov transition kernel, that is,
for any $x \in\mathsf{X}$, $A \in\mathcal{B}$, and $n \in
\mathbb{Z}^+$, $P^n(x,A) = \operatorname{Pr}(X^{(n+j)} \in A | X^{(j)} = x)$
for the Markov chain $\Phi=  \{X^{(0)}, X^{(1)}, X^{(2)}, \ldots
\}$. Let \mbox{$\|\cdot\|$} denote the total\vadjust{\goodbreak} variation norm. If the
chain is Harris ergodic, then for all $x \in\sX$ we have $\| P^n(x,
\cdot) - \varpi(\cdot) \| \to0$ as $n \to\infty$. Now suppose there
exist a real-valued function $M(x)$ on $\mathsf{X}$ and $0 < t < 1$
such that
%
\begin{equation}
\label{eqergodic} \bigl\| P^n(x, \cdot) - \varpi(\cdot) \bigr\| \leq M(x)
t^{n}.
\end{equation}
If $M$ is bounded, then $\Phi$ is \textit{uniformly ergodic} and
otherwise it is \textit{geometrically ergodic}.

A common goal of an MCMC experiment is to evaluate the quantity
$E_{\varpi}g = \int_{\mathsf{X}}{g(x) \varpi(dx)}$, where $g$ is a
real-valued function on $\sX$ whose expectation exists. Upon
simulation of the Markov chain, $E_{\varpi}g$ is approximated by the
sample average $\bar{g}_{n} =\break n^{-1} \sum_{i=0}^{n-1} g(X^{(i)})$.
This approximation is usually justified through Birkhoff's ergodic
theorem. Now, along with a moment condition on $g$, the existence of
$M$ and $t$ in (\ref{eqergodic}) ensures the existence of a central
limit theorem for the Monte Carlo error, that is, there exists $0 <
\sigma^2_g < \infty$ such that, as $n \to\infty$,
%
\begin{equation}
\label{eqclt} \sqrt{n} (\bar{g}_{n} - E_{\varpi} g)
\stackrel{d} {\to} \mathrm{N}\bigl(0, \sigma_{g}^{2}\bigr).
\end{equation}
Along with various moment conditions, the existence of $M$ and $t$ in
(\ref{eqergodic}) is also a key sufficient condition for using a
variety of methods such as batch means, spectral methods or
regenerative simulation to construct a strongly consistent estimator of
$\sigma_{g}^{2}$, ensuring asymptotically valid Monte Carlo standard
errors
(\cite{atch2011}; \cite{flegjone2010};
\cite{hobejonepresrose2002}; \cite{joneharacaffneat2006}).
Thus, when $\Phi$ is at least geometrically ergodic, a practitioner has
the tools to be as confident in their simulations as if it were
possible to make independent and identically distributed draws from
$\varpi$ (\cite{flegharajone2008}; \cite{flegjone2011}).

Much work has been done on establishing geometric and uniform
ergodicity of various versions of Metropolis--Hastings when
full-dimensional updates are used; for example, \citet{mengtwee1996}
and \citet{tier1994} studied the MHIS while
\citet{johngeye2012}, \citet{mengtwee1996},
\citet{robetwee1996}, Jarner and\break Hansen (\citeyear{jarnhans2000}),
Christensen, M{\o}ller and Waagepeter\-sen (\citeyear{chrimollwaag2001}) have established conditions under which
Metropolis yields a geometrically ergodic chain. Other research on
establishing convergence rates of Metrop\-olis--Hastings chains includes
\citet{geye1999}, \citet{jarnhans2000} and
\citet{meyntwee1994b}. Note well that none of these convergence
rate results apply to component-wise implementations of
Metropolis--Hastings. Also, a full-dimensional updating algorithm may
fail to be geometrically ergodic while many component-wise updating
samplers are. Consider the following simple example.

\begin{example} \label{extoyexampleinintroduction} Suppose $\log
\pi(x,y) = -(x^{2} + x^{2}y^{2} + y^{2})$. \citet{robetwee1996}
showed that a Me\-tropolis random walk having target $\pi$ cannot be
geometrically ergodic. Thus, one might consider com\-ponent-wise
methods where the target of each update is the relevant conditional
distribution, $X|Y=y \sim\mathrm{N}(0, \frac{1}{2}(1+y^{2})^{-1})$
or $Y|X=x \sim\mathrm{N}(0,
\frac{1}{2}(1+x^{2})^{-1})$. \citet{fortmoulroberose2003}
established geometric ergodicity of the uniform random scan
Metropolis random walk. That is, at each step one of the components
is selected with probability $1/2$ to remain fixed while a Metropolis
random walk step is performed for the other. We will show that the
Gibbs sampler is geometrically ergodic as are the random scan Gibbs
and random sequence Gibbs samplers for any selection probabilities.
\end{example}

We study conditions for ensuring geometric or uniform ergodicity for
several component-wise strategies. Despite the near ubiquity of
component-wise methods in MCMC practice, there has been very little
work on this problem. In particular, there has been almost none in
the case where the component-wise updates are done with
Metropolis--Hastings (but
see \cite{fortmoulroberose2003}; \cite{joneroberose2013}; \cite{roberose1998a}).
The one component-wise method that has received some attention in the
literature is the Gibbs sampler, especially the two-variable
deterministically updated Gibbs sampler; see, for example, the work in
\citet{dosshobe2010}, \citet{hobegeye1998},
\citet{hobejonepresrose2002}, \citet{johnjone2010},
\citet{jonehobe2004}, \citet{marchobe2004},\break
\citet{paparobe2008}, \citet{robepols1994},
\citet{roberose1999a}, \citet{romahobe2012},
Rosenthal (\citeyear{rose1995a,rose1996}), \citet{royhobe2007},
\citet{tanhobe2009} and \citet{tier1994}.

In Section~\ref{seccompwise} we fix some notation, state assumptions
and develop a general framework for compo\-nent-wise updates. Then in
Section~\ref{secconv} we study the convergence rates of
component-wise methods. In particular, we connect the convergence
rate of deterministic scan samplers with random sequence scan and
random scan methods. We also develop conditions for the uniform
ergodicity of component-wise versions of the Metropolis--Hastings
algorithm with state-independent candidate distributions. Along the
way we apply our results to two practically relevant examples,
including the Gibbs sampler for a Bayesian linear mixed model and one
involving maximum likelihood estimation for mixed models, and provide
empirical comparisons\vadjust{\goodbreak} between samplers using component-wise updates
and their full-dimen\-sional counterparts. Notably, the empirical
performance of the component-wise samplers is compel\-lingly better,
thus providing further support for the use of compo\-nent-wise methods
in practical problems; see also \citet{caffjankjone2005},
\citet{coulhoberyanholm2001}, \citet{johnjone2010},
\citet{joneharacaffneat2006}, \citet{jonehobe2001},
\citet{leejonecaffbass2013}, \citet{mccu1997} and
\citet{neat2012}. Proofs of our results and other technical material
are given in the supplement to the current article
(\cite{johnjoneneat2013b}).

\section{Component-Wise Updates}
\label{seccompwise}

Two fundamental strategies for combining Markov kernels are mixing and
composition. Suppose $P_{1},\ldots,\break P_{d}$ are Markov kernels having
common invariant distribution $\varpi$. The general composition kernel
is given by $P_{\mathrm{comp}} (x, \cdot) = (P_{1} \cdots P_{d})(x, \cdot)$. Let
$\mathbb{P}^d = \{ (r_{1},\ldots,\break r_{d}) \in\mathbb{R}^{d}
\mbox{: each } r_{i} > 0, \sum_{i=1}^{d} r_{i} = 1\}$. Define the
general mixing kernel
\[
P_{\mathrm{mix}} (x, \cdot)= r_{1} P_{1} (x, \cdot)+
\cdots+ r_{d} P_{d} (x, \cdot),\quad r \in\mathbb{P}^{d}.
\]
Then $P_{\mathrm{mix}}$ and $P_{\mathrm{comp}}$ are Markov kernels preserving the
invariance of $\varpi$ and we say that $P_{\mathrm{mix}}$ \textit{has
selection probabilities} $r$.

We can use these two strategies to create many component-wise
algorithms. Suppose $\pi$ is a density of $\varpi$ with respect to a
measure $\mu=\mu_{1} \times\cdots\times\mu_{d}$ and has support
$\mathsf{X} = \mathsf{X}_1 \times\cdots\times\mathsf{X}_d$ with
Borel $\sigma$-algebra~$\mathcal{B}$. We allow each $\mathsf{X}_{i}
\subseteq\mathbb{R}^{b_{i}}$ so that the total dimension is $b_{1} +
\cdots+ b_{d}$. If $x \in\mathsf{X}$, set $x_{(i)} = x
\setminus x_{i}$. For $i=1,\ldots, d$ let $g_{i}(y_{i} | x)$ be
a density satisfying
%
\begin{equation}
\label{eqcondition} \pi(y_{i} | x_{(i)}) = \int
g_{i}(y_{i} | x) \pi(x_{i} | x_{(i)})
\mu_{i}(dx_{i}).
\end{equation}
That is, the conditional density $\pi(x_{i} | x_{(i)})$ is invariant
for $g_{i}(y_{i} | x)$. Note that (\ref{eqcondition}) is trivially
satisfied if $g_{i}$ corresponds to an elementary Gibbs update, that
is, $g_{i}(y_{i} | x) = \pi(y_{i} | x_{(i)})$. Also, condition
(\ref{eqcondition}) is satisfied by construction if $g_{i}$
corresponds to a Metropolis--Hastings algorithm having $\pi(y_{i} |
x_{(i)})$ as its target.

Given (\ref{eqcondition}), define a Markov kernel $P_{i}$ as
%
\begin{eqnarray}
\label{eqcompwise} P_{i} (x, A) = \int_{A}
g_{i}(y_{i} | x) \delta(y_{(i)} -
x_{(i)}) \mu(dy) \nonumber\\[-8pt]\\[-8pt]
&&\eqntext{\mbox{for } A \in\mathcal{B},}
\end{eqnarray}
where $\delta$ is Dirac's delta. Then $\varpi$ is invariant for each
$P_{i}$ so that $\varpi P_{i} = \varpi$ since
\[
\int_{\mathsf{X}} \pi(x) g_{i}(y_{i} | x)
\delta(y_{(i)} - x_{(i)}) \mu(dx) = \pi(y).
\]
Since component-wise updates are not $\varpi$-irreducible, we need to
combine the $P_{i}$ in order to achieve a useful algorithm. Let $r \in
\mathbb{P}^d$ be the selection probabilities corresponding to the
components. Then we can write the random scan Markov kernel as
%
\begin{equation}
\label{eqPmix} P_{\mathrm{RS}}(x,A) = \sum_{i=1}^d
r_i P_i (x, A)
\end{equation}
and it is obvious that $\varpi P_{\mathrm{RS}} = \varpi$. Moreover, $P_{\mathrm{RS}}$
admits a Markov transition density (Mtd)
\[
h_{\mathrm{RS}}(y|x) = \sum_{i=1}^{d}
r_{i} g_{i}(y_{i} | x) \delta(y_{(i)} -
x_{(i)}).
\]
Another way to combine the $P_{i}$ is through composition, that is,
deterministically cycling through the component-wise updates one at a
time, in which case the Markov kernel is
%
\begin{equation}
\label{eqPcomp} P_{C}(x, A) = (P_{1} \cdots
P_{d}) (x, A)
\end{equation}
and it is easy to see that $\varpi P_{C} = \varpi$ and that the
associated Mtd is
\[
h_{C} (y |x) = g_{1}(y_{1} | x)
g_{2}(y_{2} | y_{1}, x_{(1)}) \cdots
g_{d}(y_{d} | y_{(d)}, x_{d}).
\]
There are $d!$ orders in which composition can be used and it is
natural to consider using mixing to combine some of them. If $r \in
\mathbb{P}^p$ for $p \le d!$, the sequence mixing kernel is given by
%
\begin{equation}
\label{eqPseqmix} P_{\RQ}(x,A) = \sum_{j=1}^{p}
r_{j} P_{C, j}(x,A),
\end{equation}
where the $P_{C,j}$ are kernels created via composition but in
different orders. Since $\varpi P_{C,j} = \varpi$ for each $j$, it is
easy to see that $\varpi P_{\RQ} = \varpi$. Clearly, the Mtd is
\[
h_{\RQ}(y|x) = \sum_{j=1}^{p}
r_{j} h_{C, j}(y|x).
\]
Note that the kernels defined in (\ref{eqPmix}), (\ref{eqPcomp})
and (\ref{eqPseqmix}) are special cases of the general definitions
of $P_{\mathrm{mix}}$ and $P_{\mathrm{comp}}$. We will employ the notation $P_{C}$,
$P_{\mathrm{RS}}$ and $P_{\RQ}$ for the special case of component-wise updates,
that is, when the $P_i$ satisfy (\ref{eqcompwise}).

\section{Convergence Rates Under Component-Wise Updates}
\label{secconv}

Most of the research on convergence rates of com\-ponent-wise MCMC
algorithms has focused on those formed by composition, such as
deterministic scan Gibbs samplers. One of our goals in this section
is to show that the convergence rates of samplers formed by mixing and
composition are related in concrete ways. However, we begin with a
brief description of some techniques for establishing the existence of
$M$ and $t$ in (\ref{eqergodic}); see Meyn and Tweedie [(\citeyear{meyntwee1993}),
Chapter 15] and \citet{roberose2004} for details and
\citet{jonehobe2001} for an accessible introduction.

Recall that $\mathcal{B}$ is the Borel $\sigma$-algebra on
$\mathsf{X}$ and $P^n(x, dy)$ denotes a $n$-step Markov transition
kernel, that is, for any $x \in\mathsf{X}$, $A \in\mathcal{B}$, and
$n \in\mathbb{Z}^+$, $P^n(x,A) = \operatorname{Pr}(X^{(n+j)} \in A |
X^{(j)} = x)$ for the Markov chain $\Phi=  \{X^{(0)}, X^{(1)},
X^{(2)}, \ldots \}$.

Suppose there exist a positive integer $n_0$, an $\epsilon> 0$, a set
$C \in\mathcal{B}$, and a probability measure $Q$ on $\mathcal{B}$
such that
%
\begin{equation}
\label{eqminorization} P^{n_0}(x, A) \geq\epsilon Q(A) \quad\mbox{for
all }x \in C, A \in\mathcal{B}.
\end{equation}
Then a \textit{minorization condition} holds on the set $C$, called a
\textit{small set}. Uniform ergodicity is equivalent to the existence
of a minorization condition on $\sX$.

Let
%
\begin{equation}
\label{eqPV} \PV(x):=E\bigl[V\bigl(X^{(t+1)}\bigr) | X^{(t)}=x
\bigr].
\end{equation}
A \textit{drift condition} holds if there exists some function $V\dvtx
\mathsf{X} \to[1,\infty)$, constants $0 < \gamma< 1$ and $k <
\infty$ and a set $C$ such that
%
\begin{equation}
\label{eqdrift}\quad \PV(x) \le\gamma V(x) + k I_{C}(x)\quad\mbox{for all
} x \in\mathsf{X}.
\end{equation}
If $C$ is small, then (\ref{eqdrift}) is equivalent to geometric
ergodicity (\cite{meyntwee1993}; Roberts and Rosenthal,
\citeyear{roberose1997c,roberose2004}).

\subsection{Uniform Ergodicity Under Mixing and Composition}
\label{secgeneral}

We begin with some results concerning samplers $P_{\mathrm{mix}}$ and
$P_{\mathrm{comp}}$, after which we will specialize to some component-wise
samplers.

\begin{theorem}
\label{thmgenmix}
If $P_{\mathrm{mix}}$ is uniformly ergodic for some selection probabilities, then
it is uniformly ergodic for all selection probabilities.
\end{theorem}

It is easy to see that if one of the component samplers is uniformly
ergodic, then $P_{\mathrm{mix}}$ will be uniformly ergodic. Also, it is
sufficient to study $P_{\mathrm{comp}}$ to establish uniform ergodicity of
$P_{\mathrm{mix}}$.

\begin{theorem}
\label{thmcomp}
Suppose $P_{\mathrm{comp}}$ is uniformly ergodic. Then the corresponding
$P_{\mathrm{mix}}$ is uniformly ergodic for any selection probabilities.
\end{theorem}

For component-wise updates it can be difficult to establish uniform
ergodicity of $P_{\mathrm{mix}}$ due to the form of its Mtd. Our next result
follows directly from Theorems~\ref{thmgenmix} and~\ref{thmcomp}
and the observation that $P_{\mathrm{RS}}$ and $P_{\RQ}$ are special cases of
$P_{\mathrm{mix}}$ and $P_{C}$ is a special case of $P_{\mathrm{comp}}$. See
\citet{laturoberose2011} and \citet{roberose1997c} for related
results.

\begin{theorem}
\label{thmsimplemixing}
If $P_{C}$ is uniformly ergodic, then $P_{\mathrm{RS}}$ and $P_{\RQ}$ are
uniformly ergodic for any selection probabilities.
\end{theorem}

\subsubsection{Component-wise independence samplers}
\label{seccwis}

It is clear that many component-wise Markov chains will not be
uniformly ergodic. For example, it is well known that Metropolis
random walks on $\mathbb{R}$ are not uniformly ergodic
(\cite{mengtwee1996}). Hence, when using such chains as the building
blocks of a component-wise algorithm one does not expect to produce a
uniformly ergodic Markov chain. On the other hand,
\citet{mengtwee1996} did show that the full-dimensional
Metropolis--Hastings independence sampler can be uniformly ergodic. We
now turn our attention to the component-wise algorithm where each
component-wise update is a Me\-tropolis--Hastings algorithm with
state-independent proposals. In this case, the composition sampler,
$P_{\mathrm{CIS}}$, the random sequence sampler, $P_{\mathrm{RQIS}}$, and the random
scan sampler, $P_{\mathrm{RSIS}}$, are all \textit{component-wise independence
samplers} (CWIS).

We are interested in establishing conditions under which the CWIS are
uniformly ergodic. By Theorem~\ref{thmsimplemixing} it is
sufficient to consider CIS. Note that since a typical $P_{\mathrm{CIS}}$
update will be some combination of accepted and rejected
component-wise proposals, the $P_{\mathrm{CIS}}$ is not truly an independence
sampler at all and, thus, the results of \citet{mengtwee1996} are not
applicable. It is, however, tempting to think that extending
Mengersen and Tweedie's (\citeyear{mengtwee1996}) work on
MHIS to $P_{\mathrm{CIS}}$ will be
straightforward. Let $p_i(\cdot)$, a density on $\mathsf{X}_i$,
denote the state-independent proposal density for the $i$th update, $i
= 1,\ldots, d$. If we let $p(x) = \prod_{i=1}^d p_i(x_i)$, a density
on $\mathsf{X}$, is the existence of $\epsilon> 0$ such that $p(x)
\geq\epsilon\pi(x) $ a sufficient condition for uniform
ergodicity of
$P_{\mathrm{CIS}}$? If we attempt to directly generalize
Mengersen and Tweedie's (\citeyear{mengtwee1996}) argument,
we are faced with $2^{d}$ cases to
consider and, hence, this approach is fruitless. However, with a
different approach we are able to give a pair of conditions that
together are sufficient for uniform ergodicity of the CWIS.

To describe our results, we require a new notation. We will continue
to let a subscript indicate the position of a vector component and a
parenthetical superscript indicate the step in a Markov chain.
Additionally, for each $i = 1, \ldots, d$, let\vspace*{2pt} $x_{[i]} = (x_1,\ldots,
x_i)$ and $x^{[i]} = (x_i,\ldots, x_d)$; let $x_{[0]}$ and
$x^{[d+1]}$ be null (vectors of dimension 0). We can now state the
result for CWIS.

\begin{theorem}
\label{thmProp1}
Consider the kernel $P_{\mathrm{CIS}}$ with proposal densities $p_i$ for $i =
1,\ldots, d$. Define $p(x):=\prod_{i=1}^d p_i(x_i)$, a density on
$\mathsf{X}$. Further suppose there exists $\delta> 0$ such that
$p(x) \geq\delta\pi(x)$ for all $x \in\mathsf{X}$, and
$\varepsilon
> 0$ such that for any $x, y \in\mathsf{X}$ with $\pi(x) > 0$ and
$\pi(y) > 0$,
%
\begin{eqnarray}
\label{eqnProp1}\qquad \pi(x) \pi(y) &=& \pi\bigl(x_{[i]}, x^{[i+1]}
\bigr) \pi\bigl(y_{[i]}, y^{[i+1]}\bigr) \nonumber\\[-8pt]\\[-8pt]
&\geq&\varepsilon\pi
\bigl(x_{[i]}, y^{[i+1]}\bigr) \pi\bigl(y_{[i]},
x^{[i+1]}\bigr)> 0\nonumber
\end{eqnarray}
for each $i = 1,\ldots, d-1$. Then, for any $x \in\mathsf{X}$ and
$A \in\mathcal{B}(\mathsf{X})$,
\[
P_{\mathrm{CIS}}(x,A) \geq\delta\varepsilon^{ \lfloor d/2 \rfloor} \pi(A)
\]
and, thus, $P_{\mathrm{CIS}}$ is uniformly ergodic. Hence, $P_{\mathrm{RQIS}}$ and
$P_{\mathrm{RSIS}}$ are also uniformly ergodic for any selection
probabilities.
\end{theorem}

The following two corollaries indicate settings where the conditions
of the theorem are easily verified.

\begin{corollary}
\label{corindependence}
Consider the kernel $P_{\mathrm{CIS}}$ with proposal densities $p_i$ for $i =
1,\ldots, d$. Define $p(x):= \prod_{i=1}^d p_i(x_i)$, a density on
$\mathsf{X}$. Further suppose there exists $\delta> 0$ such that
$p(x) \geq\delta\pi(x)$ for all $x \in\mathsf{X}$, and pairs of
positive functions $g_i$ and $h_i$ on $\mathsf{X}_i$ for $i = 1,\ldots, d$ such that
%
\begin{equation}
\label{eqdensitybounds} \prod_{i=1}^d
g_i(x_i) \leq\pi(x) \leq\prod
_{i=1}^d h_i(x_i)
\end{equation}
for any $x \in\mathsf{X}$, and $\inf_{x_i \in\mathsf{X}_i}  \{
g_i(x_i)/h_i(x_i)  \} > 0$ for each $i = 1,\ldots, d$.
Then $P_{\mathrm{CIS}}$, $P_{\mathrm{RQIS}}$ and $P_{\mathrm{RSIS}}$ are all uniformly ergodic.
\end{corollary}

The conditions of Corollary~\ref{corindependence} amount to requiring
at most
a weak form of dependence in the target distribution. The most obvious special
case is when the components of $\pi$ are jointly independent, in which case
(\ref{eqdensitybounds}) holds with equality on both sides.

\begin{corollary}
\label{corcompact}
Consider the kernel $P_{\mathrm{CIS}}$ with proposal densities $p_i$ for $i =
1,\ldots, d$. Define $p(x):= \prod_{i=1}^d p_i(x_i)$, a density on
$\mathsf{X}$. If there exist $0 < a \leq b < \infty$ and $c > 0$ such
that $a \leq\pi(x) \leq b$ and $p(x) \geq c$ for $\varpi$-almost all
$x$, then $P_{\mathrm{CIS}}$, $P_{\mathrm{RQIS}}$ and $P_{\mathrm{RSIS}}$ are all uniformly
ergodic.
\end{corollary}

\subsubsection{Maximum likelihood for mixed models}
\label{secglmm}

Let $Y_i =  \{ Y_{i1},\ldots, Y_{i m_i}  \}$ denote a vector
of observable data, and let $U_i$ denote the unobservable $i$th random
effect, for $i = 1,\ldots, k$; let $U = (U_1,\ldots, U_k)$. Assume
the $Y_i$ are independent with distribution specified conditionally on
$U = u$, so that the joint density of $Y =  \{Y_{ij}\dvtx  j= 1,\ldots, m_i; i = 1,\ldots, k  \}$ is
\[
f(y | u; \theta_1) = \prod_{i=1}^k
\prod_{j=1}^{m_i} f(y_{ij} |
u_i; \theta_1),
\]
where $\theta_1$ denotes a vector of parameters. The $U_i$ are
assumed to be independent, typically but not necessarily normally
distributed, so the joint density of $U$ is $h(u; \theta_2) =
\prod_{i=1}^k h(u_i; \theta_2)$. Then the likelihood,
\[
L(\theta; y) = \int f(y | u; \theta_1) h(u; \theta_2)
\,du
\]
is often analytically intractable so that calculating maximum
likelihood estimates and their standard errors can be
challenging. However, there are several Monte Carlo-based algorithms,
such as Monte Carlo Newton--Raphson, Monte Carlo maximum likelihood and
Monte Carlo EM, which are useful for finding maximum likelihood
estimators of the unknown parameter $\theta= (\theta_1, \theta_2)$
(\cite{caffjankjone2005}; \cite{hobe2000}; \cite{mccu1997}). A common feature is
that all three algorithms require simulation from the same target
distribution, namely, the conditional distribution of the random
effects given the data, that is, for a given value of $\theta$
\[
h(u | y; \theta) \propto f(y | u; \theta_1) h(u;
\theta_2).
\]

We consider four Markov chains having $h(u|y; \theta)$ as the
invariant density; the three component-wise independence samplers,
CIS, RQIS and RSIS having proposal densities $h(u_i; \theta_2)$ for $i
= 1,\ldots, k$ and a full-dimensional Metropolis--Hastings
Independence Sampler (MHIS) with proposals drawn from the\break marginal
distribution $h(u; \theta_2)$. To this end, the following result holds.

\begin{theorem}
\label{thmdiscretemodels}
If there exists $B(y, \theta_{1}) < \infty$ such that $f(y| u;
\theta_{1}) \le B(y, \theta_{1}) $ for all $u$, then the four Markov
chains described above, MHIS, CIS, RQIS and RSIS having $h(u|y,
\theta)$ as the invariant density, are uniformly ergodic.
\end{theorem}

We compare the empirical performance of the\break CWIS, the MHIS and a
geometrically ergodic\vadjust{\goodbreak} full-dimensional Metropolis random walk sampler
in a concrete example in Section~\ref{secbenchmark}.

\subsection{Geometric Ergodicity Under Mixing and Composition}
\label{secgeometric}

Consider $P_{\mathrm{mix}}$ and suppose each of the kernels $P_{i}$ are
geometrically ergodic in that there are nonnegative functions $M_{i} $
and $t_{i} \in(0,1)$ such that
\[
\bigl\|P_{i}(x, \cdot) - \varpi(\cdot) \bigr\| \le M_{i}(x)
t_{i}^{n}.
\]
Then the triangle inequality implies
\begin{eqnarray*}
&&
\bigl\|P_{\mathrm{mix}} (x, \cdot) - \varpi(\cdot)\bigr\|\\
&&\quad \le\bigl[r_{1}
M_{1}(x) + \cdots+ r_{d} M_{d}(x)\bigr] \bigl[
\max\{t_{1},\ldots, t_{d}\}\bigr]^{n}
\end{eqnarray*}
and, hence, we have the following observation: if each $P_{i}$ is
geometrically ergodic, then so is $P_{\mathrm{mix}}$. This demonstrates one
difference between establishing geometric and uniform ergodicity;
recall that we only required one of the $P_{i}$ to be uniformly
ergodic for $P_{\mathrm{mix}}$ to be uniformly ergodic. Also, this immediately
implies that $P_{\RQ}$ is geometrically ergodic if each of the
composition samplers are. However, it does not apply to $P_{\mathrm{RS}}$
since, in this case, each of the $P_{i}$ typically are not even
$\varpi$-irreducible. On the other hand, we have an analogue of
Theorem~\ref{thmgenmix}, albeit with an additional assumption. Recall
that a Markov kernel $P$ is \textit{reversible with respect to}
$\varpi$ if
\[
P(x, dy) \varpi(dx) = P(y, dx) \varpi(dy).
\]
%

\begin{theorem}
\label{thmreversiblemixge}
Suppose $P_{\mathrm{mix}}$ is reversible with respect to $\varpi$ for all
selection probabilities $r
\in\mathbb{P}^{d}$. If $P_{\mathrm{mix}}$ is geometrically ergodic for
some selection probability, then it is geometrically
ergodic for all selection probabilities.
\end{theorem}

Note that a special case of Theorem~\ref{thmreversiblemixge} is
that if $P_{\mathrm{RS}}$ is geometrically ergodic for some selection
probability, then it is geometrically ergodic for all selection
probabilities; see \citet{joneroberose2013} for related results.
Also, Theorem~\ref{thmreversiblemixge} does not apply to random
sequence scan samplers since these are not reversible for all selection
probabilities.

\subsubsection{Two-variable settings}
\label{sectwovar}

We consider the case where $\varpi$ has a density $\pi(x,y)$ with
respect to $\mu_1\times\mu_2$ and has support $\sX_{1} \times\sX_{2}
\subseteq\mathbb{R}^{b_{1}} \times\mathbb{R}^{b_{2}}$. Let
$\pi_{X|Y}(x|y)$ and $\pi_{Y|X}(y|x)$ be the full conditional
densities and $\pi_{X}$ and $\pi_{Y}$ be the marginal densities
derived from $\pi$ ($\varpi_X$ and $\varpi_Y$ are the marginal
distributions). This setting, though less general than that of the
previous section, has many practical applications. For instance, it
is the foundation for data augmentation methods
(\cite{hobe2011}; \cite{tannwong1987}) and many MCMC methods for practically
relevant statistical models
(\cite{johnjone2010}; \cite{romahobe2012}; \cite{royhobe2007}). We will consider
two settings here: specifically, we begin with the case where sampling from
$\pi_{X|Y}$ and $\pi_{Y|X}$ is possible and later turn our attention
to the case where one of the Gibbs updates is replaced by a
Metropolis--Hastings update.

When sampling from $\pi_{X|Y}$ and $\pi_{Y|X}$ is easy the Markov
kernel formed by composition, say, $P_{\mathrm{GS}}$, is the usual Gibbs sampler (GS)
having Mtd
%
\begin{equation}
\label{eqDUGSMtd} h_{\mathrm{GS}}\bigl(x', y'
| x, y\bigr) = \pi_{X|Y}\bigl(x' | y\bigr)
\pi_{Y|X}\bigl(y'|x'\bigr).
\end{equation}
Of course, the other update order is also a Gibbs sampler, denoted
$\tilde{P}_{\mathrm{GS}}$. Also, each of the marginal sequences $\{X^{(n)}\}$ and
$\{Y^{(n)}\}$ have one-step Markov kernels $P_{X}$ and $P_{Y}$ with
Mtds
\[
h_{X}\bigl(x' | x\bigr) = \int\pi_{X|Y}
\bigl(x' | y\bigr) \pi_{Y|X}(y|x) \mu_2(dy)
\]
and
%
\begin{equation}
\label{eqGibbsY}\quad h_{Y}\bigl(y'|y\bigr) = \int
\pi_{Y|X}\bigl(y' | x\bigr) \pi_{X|Y}(x|y)
\mu_1(dx),
\end{equation}
respectively. Moreover, it is easy to see that $P_{Y}^{m}$ admits an
$m$-step Mtd $h_{Y}^{m}(y'|y)$ as do $P_{X}^{m}$, $P_{\mathrm{GS}}^{m}$ and
$\tilde{P}_{\mathrm{GS}}^{m}$. Note that $\pi_{X}$ is invariant for
$\{X^{(n)}\}$ and $\pi_{Y}$ is invariant for $\{Y^{(n)}\}$.

It is well known that $P_{X}$, $P_{Y}$, $P_{\mathrm{GS}}$ and $\tilde{P}_{\mathrm{GS}}$
all converge at the same qualitative rate
(\cite{diacetal2008}; \cite{robe1995}; \cite{roberose2001}). In particular, if
one is geometrically ergodic, then so are the others. This
relationship has been routinely exploited in the analysis of Gibbs
samplers for practically relevant statistical models, where it is
often easier to analyze $P_{Y}$ or $P_{X}$ than $P_{\mathrm{GS}}$. Putting
these observations together with our above work says that if one of
$P_{X}$, $P_{Y}$, $P_{\mathrm{GS}}$ or $\tilde{P}_{\mathrm{GS}}$ are geometrically
ergodic, then so are the others and so is the random sequence Gibbs
sampler~$P_{\mathrm{RQGS}}$.

The first result of this subsection connects the convergence rate of
$P_{X}$, $P_{Y}$, $P_{\mathrm{GS}}$ and $\tilde{P}_{\mathrm{GS}}$ to the random scan
Gibbs sampler $P_{\mathrm{RSGS}}$. Note that by using (\ref{eqPV}) and
(\ref{eqGibbsY}) we have that
\[
P_{Y} W(y) = \int_{\sX_{2}} W\bigl(y'
\bigr) h_{Y}\bigl(y'|y\bigr) \mu_{2}
\bigl(dy'\bigr).
\]
%

\begin{theorem}
\label{thmRQGSandRSGS}
Suppose there exists $\lambda< 1$, $W\dvtx\break  \sX_2 \rightarrow
\mathbb{R}^{+}$, $b< \infty$ such that
%
\begin{equation}
\label{eqYdrift} P_{Y} W(y) \leq\lambda W(y) + b.\vadjust{\goodbreak}
\end{equation}
Let $C_{d} = \{ y\dvtx  W(y) \le d\}$ and suppose there is a $g\dvtx\break  \sX_{2}
\to\mathbb{R}^{+}$ and a $d_{0} >0$ such that for some $m
\ge1$
%
\begin{equation}
\label{eqmstepmincon}\qquad h_{Y}^{m} \bigl(y' |
y\bigr) \ge g\bigl(y'\bigr) \quad\mbox{for all } y \in C_{d}
\mbox{ and } d \ge d_{0}.
\end{equation}
Then $P_{Y}$, $P_{X}$, $P_{\mathrm{GS}}$ and $\tilde{P}_{\mathrm{GS}}$ are geometrically
ergodic as are $P_{\mathrm{RQGS}}$ and $P_{\mathrm{RSGS}}$.
\end{theorem}

There is a simple sufficient condition for (\ref{eqmstepmincon});
suppose there is a $l\dvtx  \sX_{2} \to\mathbb{R}^{+}$ such that
$\pi_{X|Y}(x | y) \ge l(x)$ for all $(x,y) \in\sX_{1} \times C_{d}$
with $d \ge d_{0}$. Then if $y \in C_{d}$,
\begin{eqnarray*}
h_{Y}\bigl(y'|y\bigr) &=& \int_{\sX_{1}}
\pi_{Y|X}\bigl(y'|z\bigr) \pi_{X|Y}(z|y)
\mu_{1}(dz) \\
&\ge&\int_{\sX_{1}} \pi_{Y|X}
\bigl(y'|z\bigr) l(z) \mu_{1}(dz) = g
\bigl(y'\bigr).
\end{eqnarray*}
Although we do not state it formally, it is clear from our proof that
the same conclusions obtain if we were to reformulate the conditions
in terms of $P_{X}$ instead of~$P_{Y}$.

\begin{example}
Consider the Gibbs sampler defined in Example~\ref{extoyexampleinintroduction}. If $W(y) = y^{2}$, it is easy to see that $P_{Y} W
(y) \le\lambda W(y) + 0.5$ for any $0 < \lambda< 1$. Moreover,
(\ref{eqmstepmincon}) holds since if $y \in C_{d}$ for any $d > 0$,
then $\pi_{X|Y}(x|y) \ge\pi^{-0.5} e^{-(1+d) x^{2}}$. Hence, the
claims of Example~\ref{extoyexampleinintroduction} hold by the
theorem.
\end{example}

Of the Gibbs samplers for practically relevant statistical problems
proved to be geometrically ergodic---see the references in
Section~\ref{secIntro}---only those in \citet{dosshobe2010} and
\citet{hobegeye1998} have had more than two components. Thus,
Theorem~\ref{thmRQGSandRSGS} can be coupled with existing results
to obtain the geometric ergodicity of the random sequence and random
scan versions of many Gibbs samplers which have been proved
geometrically ergodic.

Now suppose we are able to draw from $\pi_{X|Y}$, but instead of
sampling from $\pi_{Y|X}$ we substitute a Me\-tropolis--Hastings step
$g_{2}$ having proposal density~$p_{2}$. This results in a hybrid
composition sampler (often called Metropolis--Hastings-within-Gibbs)
having Markov kernel $P_{\mathrm{HC}}$ and Mtd
\[
h_{\mathrm{HC}} \bigl(x', y' | x, y\bigr) =
\pi_{X|Y}\bigl(x'|y\bigr) g_{2}
\bigl(y' | x', y\bigr).
\]
Then the marginal $Y$-sequence is Markovian with kernel $P_{Y}$ having Mtd
%
\begin{equation}
\label{eqCMHY}\quad h_{Y}\bigl(y'|y\bigr)= \int
\pi_{X|Y}(x|y) g_{2}\bigl(y' | x, y\bigr)
\mu_1(dx)
\end{equation}
with invariant density $\pi_{Y}$ but the marginal $X$-se\-quence is not
Markovian. Nevertheless, Robert [(\citeyear{robe1995}), Theorem 4.1] showed that
the $X$- and $Y$-sequences\vadjust{\goodbreak} converge at the same rate in total
variation norm. That is, let $\tilde{P}_X^{n}((x,y), \cdot)$ be the marginal
distribution of $X^{(n)}$ given initial state $(X^{(0)}, Y^{(0)}) = (x,y)$,
then for each $n \ge1$
\begin{eqnarray*}
\bigl\|P_{Y}^{n+1}(y, \cdot) - \varpi_{Y}(\cdot) \bigr\|
&\le& \bigl\|\tilde{P}_X^{n}\bigl((x,y),\cdot\bigr) -
\varpi_{X}(\cdot) \bigr\|\\
&\le&\bigl\|P_{Y}^{n}(y,\cdot) -
\varpi_{Y}(\cdot) \bigr\|.
\end{eqnarray*}
An easy calculation shows that
\[
\bigl\|P_{Y}^{n+1} (y, \cdot) - \varpi_{Y}(\cdot) \bigr\|
\le \bigl\|P_{\mathrm{HC}}^{n} \bigl((x,y), \cdot\bigr)- \varpi(\cdot)\bigr\|.
\]
It is also easy to see that the $Y$-sequence is de-initializing for
$P_{\mathrm{HC}}$ (\cite{roberose2001}). Hence, $P_{Y}$ is geometrically
ergodic if
and only if $P_{\mathrm{HC}}$ is geometrically ergodic. Our next result
connects the convergence rate of $P_{Y}$ and $P_{\mathrm{HC}}$ to the
convergence rate of the random scan hybrid chain having kernel
$P_{\mathrm{RSH}}$ and Mtd
\begin{eqnarray*}
h_{\mathrm{RSH}}\bigl(x', y' | x, y\bigr) &=& r
\pi_{X|Y}\bigl(x'|y\bigr) \delta\bigl(y'-y
\bigr) \\
&&{}+ (1-r) g_{2}\bigl(y'|x,y\bigr) \delta
\bigl(x'-x\bigr).
\end{eqnarray*}
It is straightforward to show that $P_{\mathrm{RSH}}$ is
reversible with respect to $\varpi$. Note that by using (\ref{eqPV}) and
(\ref{eqCMHY}) we have that
\[
P_{Y} W(y) = \int_{\sX_{2}} W\bigl(y'
\bigr) h_{Y}\bigl(y'|y\bigr) \mu_{2}
\bigl(dy'\bigr).
\]
%

\begin{theorem}
\label{thmHybrid}
Suppose there exists $W\dvtx \sX_2 \to\mathbb{R}^{+}$ and
constants $\lambda$, $b< \infty$ such that
%
\begin{equation}
\label{eqHCYdrift} P_{Y} W(y) \leq\lambda W(y) + b.
\end{equation}
Let $C_{d} = \{ y\dvtx  W(y) \le d\}$ and suppose there is a $g\dvtx\break  \sX_{2}
\to\mathbb{R}^{+}$ and a $d_{0} >0$ such that for some $m
\ge1$
%
\begin{equation}
\label{eqHCmstepmincon}\qquad h_{Y}^{m} \bigl(y'
| y\bigr) \ge g\bigl(y'\bigr) \quad\mbox{for all } y \in
C_{d} \mbox{ and } d \ge d_{0}.
\end{equation}
Then $P_{Y}$ and $P_{\mathrm{HC}}$ are geometrically ergodic. Further suppose
the proposal density $p_{2}$ for the Metropolis--Hastings step $g_{2}$
satisfies either:
\begin{enumerate}
\item$p_{2}(z|x,y)=p_{2}(y|x,z)$ and there exists $K < \infty$\break such
that $p_2(z |x, y) / p_2(z | x, u) \le K$, or
\item$p_{2}(z | x, y)=p_{2}(z|x)$.
\end{enumerate}
Then $P_{\mathrm{RSH}}$ is geometrically ergodic.
\end{theorem}

As with the Gibbs sampler setting, there is a simple sufficient
condition for (\ref{eqHCmstepmincon}); suppose there is a
nonnegative function $l$ such that for $y \in C_{d}$
\[
\pi_{X|Y}(x|y)g_{2}(z|x,y) \ge l (x, z).
\]
In this case, if $y \in C_{d}$, then
\begin{eqnarray*}
h_{Y}\bigl(y'|y\bigr) &=& \int_{\sX_{1}}
\pi_{Y|X}\bigl(y'|x\bigr) g_{2}
\bigl(y'|x,y\bigr) \mu_{1}(dx) \\
&\ge&\int
_{\sX_{1}}l\bigl(x,y'\bigr) \mu_{1}(dx) =
g\bigl(y'\bigr).
\end{eqnarray*}

\subsubsection{Gibbs samplers for a Bayesian linear mixed model}
\label{secBLME}

Let $Y$ denote an $N \times1$ response vector and let $\beta$ be a $p
\times1$ vector of regression coefficients, $u$ be a $k \times1$
vector of random effects, $X$ be a known $N \times p$ design matrix
and $Z$ be a known $N \times k$ matrix. Then for $r,s,t \in
\{1,2,\ldots\}$ suppose
\begin{eqnarray*}
Y|\beta, u, \lambda_{R}, \lambda_{D} & \sim &
\mathrm{N}_{N}\bigl(X\beta+ Zu, \lambda_{R}^{-1}
I_{N}\bigr),
\\
\beta| u, \lambda_{R}, \lambda_{D} &\sim&\sum
_{i=1}^{r} \eta_{i} \mathrm{N}_{p}
\bigl(b_{i}, B^{-1}\bigr),\\
u | \lambda_{R},
\lambda_{D} &\sim&\mathrm{N}_{k}\bigl(0,
\lambda_{D}^{-1} I_{k}\bigr),
\\
\lambda_{R} &\sim&\sum_{j=1}^{s}
\phi_{j} \operatorname{Gamma}(r_{j1}, r_{j2}),\\
\lambda_{D} &\sim&\sum_{l=1}^{t}
\psi_{l} \operatorname{Gamma}(d_{l1}, d_{l2}),
\end{eqnarray*}
where the mixture parameters $\eta_{i}$, $\phi_{j}$ and $\psi_{l}$
are known nonnegative constants satisfying
\[
\sum_{i=1}^{r} \eta_{i} = \sum
_{j=1}^{s} \phi_{j} = \sum
_{l=1}^{t} \psi_{l}= 1.
\]
Note that we say $W \sim\operatorname{Gamma}(a,b)$ if it has density
proportional to $w^{a-1} e^{-bw} I(w > 0)$. Finally, we also assume
$X^{T} Z = 0$, $b_{i} \in\mathbb{R}$, and the positive definite
matrix $B$ are known and the hyperparameters $r_{j1}, r_{j2}, d_{l1}$ and
$d_{l2}$ are positive.

Let $\xi=(u^{T}, \beta^{T})^{T}$ and $\lambda=(\lambda_{R},
\lambda_{D})^{T}$. Then the posterior density is characterized by
\[
\pi(\xi, \lambda| y) \propto f(y|\xi, \lambda) f(\xi|\lambda) f(\lambda),
\]
where $y$ is the observed data and $f$ denotes a generic density. It
is straightforward to derive the conditional distributions of $\xi
|\lambda,y$ and
$\lambda|\xi, y$, which are reported here. Let
\[
v_{1}(\xi) = (y - X\beta- Zu)^{T}(y-X\beta- Zu)
\]
and
\[
v_{2}(\xi) = u^{T}u.
\]
Then the distribution of $\lambda| \xi, y$ has density
\[
f(\lambda| \xi, y) = \sum_{j=1}^{s} \sum
_{l=1}^{t} \phi_{j}
\psi_{l} f_{1j}(\lambda_{R}|\xi, y)
f_{2l}(\lambda_{D} | \xi, y),
\]
where $f_{1j}(\cdot| \xi, y)$ is a Gamma$(r_{j1} + N/2, r_{j2} +
v_{1}(\xi) /2)$ density and $f_{2l}(\cdot| \xi, y)$ denotes a
$\operatorname{Gamma}(d_{l1} + k/2$, $d_{l2} + v_{2}(\xi)/2)$ density. Next, $\xi|
\lambda, y \sim\sum_{i=1}^{r} \eta_{i} \mathrm{N}(m_0,\break \Sigma^{-1})$,
where
\[
\Sigma^{-1} = \pmatrix{ \bigl(\lambda_{R} Z^{T}
Z + \lambda_{D}I_{k}\bigr)^{-1} & 0
\cr
0 &
\bigl(\lambda_{R} X^{T}X + B\bigr)^{-1}}
\]
and
\[
m_0 = \pmatrix{ \lambda_{R}\bigl(\lambda_{R}
Z^{T} Z + \lambda_{D}I_{k}\bigr)^{-1}
Z^{T} y
\cr
\bigl(\lambda_{R} X^{T}X + B
\bigr)^{-1}\bigl(\lambda_{R} X^{T} y + B b
\bigr)}.
\]
It is straightforward to implement any of the Gibbs sampling
strategies. For example, consider the Gibbs sampler that
updates $\xi$ followed by $\lambda$ having Mtd [recall (\ref{eqDUGSMtd})]
\[
h_{\mathrm{GS}}\bigl(\xi', \lambda' |\xi,
\lambda\bigr) = f\bigl(\xi' | \lambda, y\bigr) f\bigl(
\lambda'|\xi', y\bigr).
\]
We can similarly use the full conditionals and the recipes described
earlier to construct Mtds for the related Markov chains, say,
$h_{\xi}$, $h_{\lambda}$, $\tilde{h}_{\mathrm{GS}}$, $h_{\mathrm{RSGS}}$\break and $h_{\mathrm{RQGS}}$.

\citet{johnjone2010} establish (\ref{eqYdrift}) and
(\ref{eqmstepmincon}) for the marginal $\xi$-sequence having Mtd
$h_{\xi}$ and, hence, we can appeal to Theorem~\ref{thmRQGSandRSGS}
to establish the geometric ergodicity of the Gibbs samplers. Let
$x_{i}$ and $z_{i}$ be the $i$th rows of $X$ and $Z$, respectively.
Define
\begin{eqnarray*}
G_{i}(\lambda) &=& \sum_{m=1}^{N}
\bigl[ E_{i} (y_{m} - x_{m} \beta-
z_{m} u | \lambda, y)\bigr]^{2}\\
&&{} + \sum
_{m=1}^{k} \bigl[ E_{i} (u_{m}
| \lambda, y)\bigr]^{2},
\end{eqnarray*}
where $E_{i}$ denotes expectation with respect to the $N_{k+p}(m_{i},
\Sigma^{-1})$ distribution.
%
\begin{theorem}
\label{thmBLME}
Assume there exists some $K<\infty$ such that $G_{i}(\lambda) \le K$.
If for all $j \in\{1,\ldots,s\}$ and $l \in\{1,\ldots,t\}$
\[
r_{j1} > 0 \vee\frac{1}{2} \Biggl[\sum
_{i=1}^N z_i\bigl(Z^TZ
\bigr)^{-1}z_i^T - N + 2 \Biggr]
\]
and
\[
d_{l1} > 1,
\]
then the marginal $\xi$- and $\lambda$-chains and GS are geometrically
ergodic as are RSGS and RQGS.
\end{theorem}

\citet{johnjone2010} provide other conditions under which GS is
geometrically ergodic and, hence, the theorem does not exhaust the
conditions under which RQGS and RSGS are geometrically ergodic; see
also \citet{roma2012} for some improvements on the results of
\citet{johnjone2010}. In Section~\ref{seclme} we consider a
special case of\vadjust{\goodbreak} our model and provide an empirical comparison of GS,
RQGS and RSGS with a full-dimensional Metropolis sampler.

\section{Examples}
\label{secExamples}

We consider two examples based on the settings introduced in
Sections~\ref{secglmm} and~\ref{secBLME}. In each case we consider
the finite sample empirical performance of some component-wise MCMC
algorithms against full-dimensional updates. This comparison is based
on several measures of efficiency, which are now described.

If $E_{\varpi} |g(X)|^{2+\delta} < \infty$ for some $\delta>0$ and the
Markov chain is geometrically ergodic, then a central limit theorem,
recall (\ref{eqclt}), holds. Therefore, $t_*\sigma_{g}/\sqrt{n}$
gives the half-width of an asymptotically valid confidence interval
for $E_{\varpi} g$ where $t_*$ is an appropriate quantile. The width
of the interval can be used to determine the number of iterations
required to achieve some desired level of precision
(\cite{flegharajone2008}; \cite{flegjone2011}; \cite{joneharacaffneat2006}).
We might also measure Markov chain efficiency relative to the
efficiency of a would-be random sample from $\varpi$. One such
measure, the integrated autocorrelation time (ACT)
\[
\mathrm{ACT} = \frac{\sigma^{2}_{g}}{\operatorname{Var}_{\varpi}(g(X))},
\]
compares the variability of the Monte Carlo estimate to that of an
estimate based on a random sample of the same size. In practice,
$\operatorname{Var}_{\varpi}(g(X))$ and $\sigma^{2}_{g}$ are
unknown. However,
a consistent estimator of $\operatorname{Var}_{\varpi}(g(X))$ is
given by the
sample variance, $\widehat{\operatorname{Var}}_{\varpi}(g(X))$, and,
because the
chains are geometrically ergodic, the consistent batch means
estimator of \citet{joneharacaffneat2006}, say, $\hat{\sigma}^{2}_{g}$,
provides a consistent estimator of $\sigma^{2}_{g}$.

For a given sample size, the quality of Monte Carlo estimates can be
assessed using the mean squared error (MSE). To estimate the MSE, we run
$m$ independent replications of each chain, each of which is of length
$n$, producing independent estimates $\bar{g}^{(1)}_{n},\ldots,\break
\bar{g}^{(m)}_{n}$ and an independent estimate based on a long run of
a given chain, say, $\bar{g}^{*}$. The estimated MSE is
\[
\widehat{\mathrm{MSE}}_{m} (\bar{g}_{n}) = \frac{1}{m} \sum
_{i=1}^{m} \bigl( \bar{g}^{(i)}_{n}
- \bar{g}^{*}\bigr)^{2}.
\]

The above quantities allow examination of efficiency only in terms of
estimating $\mathrm{E}_{\varpi} g(X)$. We also compare how the chains
move around the state space using the expected square Euclidean jump
distance (ESEJD), that is, the expected squared distance\vadjust{\goodbreak} between
successive draws of the Markov chain $X^{(i)}$ and $X^{(i+1)}$. If \mbox{$\|
\cdot\|_{2}$} denotes the standard Euclidean norm, then ESEJD is the
expected value of the mean square Euclidean jump Distance (MSEJD) at
stationarity, where for a chain of length $n$,
\[
\mathrm{MSEJD}:= \frac{1}{n-1}\sum_{i=1}^{n-1}
\bigl\| X^{(i+1)} - X^{(i)}\bigr\|^2_2.
\]
Given $m$ independent replications of each chain, each of which is
length $n$,
we can estimate ESEJD with
\[
\widehat{\mathrm{ESEJD}}_{m} = \frac{1}{m} \sum
_{i=1}^{m}\operatorname{MSEJD}^{(i)}.
\]
In addition to the above numerical summaries, we include standard
graphical summaries such as trace plots. Taken together, these
measures give us a reasonable picture of the empirical performance of
the various algorithms examined below.

\subsection{A Bayesian Linear Mixed Model}
\label{seclme}

Consider a Bayesian version of the balanced random intercept model for
$k$ subjects\vspace*{1pt} and $m \ge2$ observations per subject. Let
$y_i=(y_{i1},\ldots,y_{im})^T$ be the\vspace*{1pt} data for subject $i$ and
$Y=(y_1^T,\ldots,y_k^T)^T$ denote the overall $N\times1$ response
vector where $N = km$. Further, let $u=(u_1,\ldots,u_k)^T$ be a
vector of subject effects and $X$ be a full column rank $N\times p$
design matrix corresponding to $\beta$, a $p \times1$ vector of
regression coefficients. Then the first level of the hierarchy is
\[
Y | \beta, u, \lambda_R, \lambda_D \sim
\mathrm{N}_N\bigl(X\beta+ Zu, \lambda_R^{-1}I_N
\bigr)
\]
for $Z = I_k \otimes1_m$, where $\otimes$ denotes the Kronecker
product and $1_m$ is an $m \times1$ vector of ones. At the next
stage,
\[
\beta|\lambda_R,\lambda_D \sim\mathrm{N}_p
\bigl(b, B^{-1}\bigr)
\]
and
\[
u |\lambda_R,\lambda_D \sim\mathrm{N}_k\bigl(0,
\lambda_D^{-1}I_k\bigr)
\]
for known $b \in\mathbb{R}^p$ and positive definite matrix $B$. Finally,
\[
\lambda_R \sim\operatorname{Gamma}(r_1,r_2)
\quad\mbox{and}\quad \lambda_D \sim
\operatorname{Gamma}(d_1,d_2),
\]
where $r_1,r_2,d_1,d_2$ are positive. This hierarchy is a special case
of the
Bayesian general linear model of Section~\ref{secBLME} and it follows from
Theorem~\ref{thmBLME} that if $d_{1} > 1$, then GS, RQGS and
RSGS are geometrically ergodic.

We present an empirical comparison of the GS, uniform RQGS and uniform
RSGS algorithms. We also\vadjust{\goodbreak} compare the three Gibbs samplers to a
full-dimensional Metropolis random walk. In our comparison we focus
on estimating the posterior expectation of $\beta$, that is,
$\mathrm{E}(\beta|y)$.

Our Metropolis random walk (RW) uses a multivariate Normal proposal
distribution centered at the current value of the chain and with a
diagonal covariance matrix. We set the diagonal elements equal to
those of $\hat{\Sigma}^2$, where $\hat{\Sigma}$ is an estimate of the
posterior covariance matrix obtained from an independent run of $10^5$
iterations of the GS. For the settings described below, our
RW has a proposal acceptance rate of approximately 0.30. We do not
know if this RW Markov chain is geometrically ergodic.

We simulated data (values of $y$) under the following settings.
Set $k=10$, $m=5$, and $p=1$, and $X=(x_1^T,\ldots,x_{10}^T)^T$, where
for all $i$, $x_i^T = (-0.50,-0.25,0,\break0.25,0.50)$ with $b=0$,
$B^{-1}=0.1$, and $r_1=r_2=d_1=d_2=2$. Assuming the true nature of
this data is unknown, we simulate the four Markov chains under the
hyperparameter setting with $b=0$, $B^{-1} = 0.1$, and $r_1= r_2= d_1=
d_2= 3$. Finally, all chains are started from the prior means,
$(\beta^{(0)},u^{(0)},\lambda_R^{(0)},\lambda_D^{(0)}) =
(0,0_{k},1,1)$ where $0_{k}$ is a $k \times1$ vector of zeroes.

Since $E[\beta^{4} | y] < \infty$, the geometric ergodicity of GS,
RQGS and RSGS guarantees a central limit theorem for the Monte Carlo
error $\bar{\beta}_{n} - \mathrm{E}(\beta| y)$ with the variance
of the asymptotic distribution denoted $\sigma^{2}_{\beta}$.

\begin{figure*}[t!]

\includegraphics{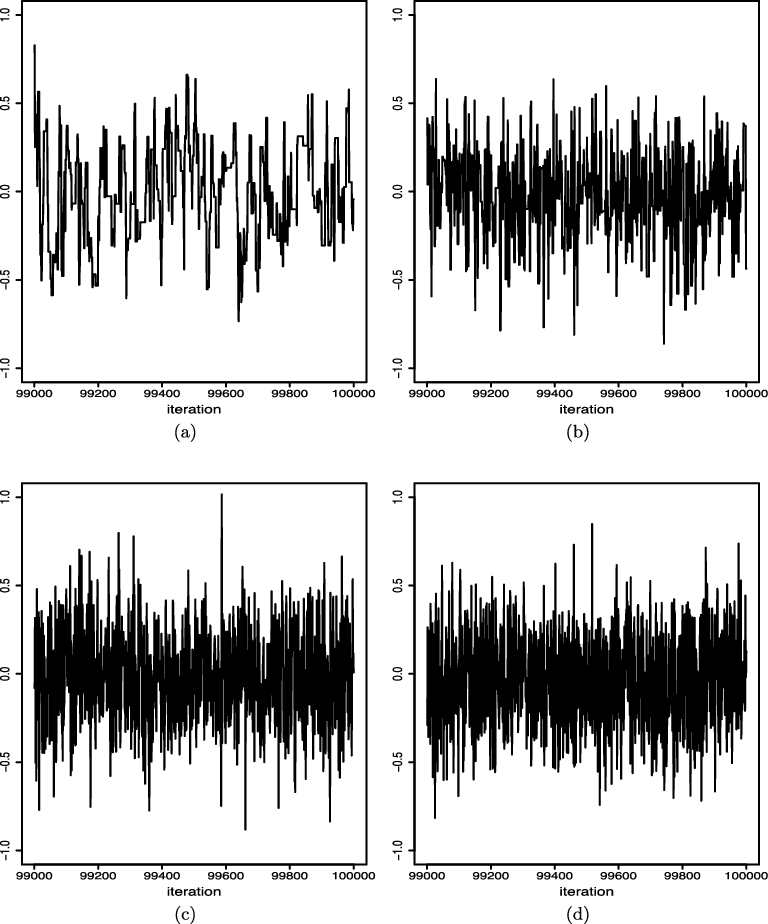}

\caption{Trace plots of $\beta$ for iterations 9.9e4 through 1e5 of
the \textup{(a)} RW, \textup{(b)} RSGS, \textup{(c)} RQGS and \textup{(d)} GS in the Bayesian linear mixed
model example of Section \protect\ref{seclme}.}
\label{figtrace}
\end{figure*}

We ran each algorithm (RW, RSGS, RQGS and GS) independently for
$10^5$ iterations. Trace plots of the final 1000
$\beta$ iterations are shown in Figure~\ref{figtrace}. Mixing
appears to
be substantially quicker for the Gibbs samplers than for the RW, while RQGS
and GS appear to be more efficient than the RSGS.

\begin{table*}[t!]
\caption{Results for the Bayesian linear mixed model example of
Section~\protect\ref{seclme}. Estimates of $\mathrm{E}(\beta|y)$,
$\sigma^2_{\beta}$
and $\operatorname{Var}(\beta|y)$ are based on $n=10^5$; middle
columns show
half-width of 95\% confidence interval ($t_* = 1.960$) and integrated
autocorrelation time (ACT). MSE Ratios are relative to GS, with
standard errors given in parentheses. Final column shows estimated
ESEJD, with standard error in parentheses}
\label{tabex1}
\begin{tabular*}{\tablewidth}{@{\extracolsep{\fill}}ld{2.3}ccd{2.3}cc@{}}
\hline
\textbf{Algorithm} & \multicolumn{1}{c}{$\bolds{\bar{\beta}_n}$}
& \multicolumn{1}{c}{$\bolds{\hat{\sigma}^2_\beta}$} &
\multicolumn{1}{c}{$\bolds{t_*\hat{\sigma}_\beta/\sqrt{n}}$}
& \multicolumn{1}{c}{\textbf{ACT}} &
\multicolumn{1}{c}{\textbf{MSE ratio}} &
\multicolumn{1}{c@{}}{$\bolds{\widehat{\mathrm{ESEJD}}}$} \\
\hline
RW & -0.018 & 0.794 & 0.0055 & 10.919
& 5.55 (0.32) & 0.26 (0.0002) \\
RSGS & -0.016 & 0.243 & 0.0031 & 3.375
& 2.07 (0.13) & 3.20 (0.0014) \\
RQGS & -0.016 & 0.071 & 0.0017 & 0.986
& 0.98 (0.05) & 6.19 (0.0013) \\
GS & -0.015 & 0.083 & 0.0018 & 1.153
& 1.00 (0.00) & 6.19 (0.0012) \\
\hline
\end{tabular*}
\end{table*}

The differences in the trace plots between the four simulations are
reflected in the interval half-width and ACT estimates given in
Table~\ref{tabex1}. For equivalent sample sizes, the RW half-width
is nearly two times that of the RSGS and approximately three times as large
as those of the GS and RQGS. In addition, the ACTs indicate that
nearly eleven RW samples and more than three RSGS samples are
required for each random draw from $\pi$ in order to achieve the same
level of precision for estimates of $\mathrm{E}(\beta|y)$. On the other
hand, each RQGS sample and GS sample is approximately as effective as
a random draw.

In order to estimate the MSE of the Monte Carlo estimates, we simulated
$m=10^{3}$ independent replications of RW, RSGS, RQGS and GS for
$n=10^4$ iterations each and took $\bar{\beta}^{*}$ to be an estimate
of $\mathrm{E}(\beta|y)$ obtained from $10^5$ iterations of the RW
chain. The estimated MSE ratios relative to the GS,
\[
\frac{\widehat{\operatorname{MSE}}(\bar{\beta
}_{n,*})}{\widehat{\operatorname{MSE}}
(\bar{\beta}_{n,\mathrm{GS}})}
\]
are also given in Table~\ref{tabex1} along with standard errors.
Notice that ratios greater than one favor GS. Hence, these results are
consistent with those above, which suggest that the single block update
RW is less efficient than the Gibbs samplers with respect to
estimation of $\mathrm{E}(\beta|y)$.

Estimation of the ESEJD is based on the same $m=1000$ independent
replications of RW, RSGS, RQGS and GS for $n=10^4$ iterations each.
The estimates are reported along with standard errors in
Table~\ref{tabex1}. The message here is consistent with the above
discussions. The RW appears to be less efficient than the Gibbs
samplers in exploring the support of the posterior. Among the Gibbs
samplers, there is little difference in the performance quality of the
RQGS and GS, whereas both are more efficient than the RSGS.

\subsection{A Logit-Normal Mixed Model}
\label{secbenchmark}

Consider the following special case of the mixed model defined in
Section~\ref{secglmm}. Suppose, conditional on $U = u$, the
observations $Y_{ij}$ are independently distributed as
$\operatorname{Bernoulli}(p_{ij})$, where $\operatorname{logit}(p_{ij}) =\break \beta
x_{ij} + u_i$ for $j = 1,\ldots, m_i$ and $i = 1,\ldots, k$, where
the $x_{ij}$ are covariates. Let the random effects $U_1,\ldots,
U_k$ be i.i.d. $\operatorname{Normal}(0, \sigma^2)$. 
With the parameters $\theta= (\beta, \sigma^2)$ treated as fixed,
the target density is
%
\begin{eqnarray}
\label{eqbenchmarktarget}
&&h(u | y; \theta) \nonumber\\
&&\quad\propto\exp \Biggl\{ \sum
_{i=1}^k \Biggl[ u_i y_{i+}
- \sum_{j=1}^{m_i} \log \bigl( 1 +
e^{ \beta x_{ij} + u_i
} \bigr) \\
&&\hspace*{141.5pt}\qquad{}- \frac{ u_i^2 }{2 \sigma^2} \Biggr] \Biggr\},\nonumber
\end{eqnarray}
where $y_{i+} = \sum_{j=1}^{m_i} y_{ij}$ for $i = 1,\ldots, k$.

In Section~\ref{secglmm} we introduced the MHIS, CIS, RQIS and RSIS
algorithms. In the current context the conditions of
Theorem~\ref{thmdiscretemodels} are satisfied and, hence, those four
samplers are uniformly ergodic. In addition, we consider a
full-dimensional Metropolis random walk (RW) sampler with normally
distributed jump proposals, that is, the proposal density is
%
\begin{equation}
\label{eqMRWprop} p\bigl(u, u^*\bigr) \propto\exp \biggl\{ -\frac{1}{2 \tau^2}
\bigl\|u^* - u\bigr\|^{2}_{2} \biggr\},
\end{equation}
where \mbox{$\| \cdot\|_{2}$} denotes the standard Euclidean norm and $\tau^{2}$
is a tuning parameter.
Then the following result holds.

\begin{theorem}
\label{thmlogitnormal}
The full-dimensional Metropolis random walk sampler with invariant density
(\ref{eqbenchmarktarget}) and proposal
density (\ref{eqMRWprop}) is geometrically ergodic.
\end{theorem}

We compare the empirical performance of the algorithms in the
context of implementing a Monte Carlo EM (MCMC) algorithm. Now at
each step the MCEM requires a Monte Carlo approximation to
the so-called $Q$-function
\[
Q(\theta; \tilde{\theta}) = \int{ l_c(\theta; y, u) h(u | y;
\tilde{\theta}) \,du},
\]
where
\begin{eqnarray*}
&&l_c(\theta; y, u) \\
&&\quad= \sum_{i=1}^k
\sum_{j=1}^{m_i} \bigl[ y_{ij} (
\beta x_{ij} + u_i) - \log \bigl( 1 + e^{\beta x_{ij} + u_i}
\bigr) \bigr] \\
&&\qquad{}- \frac{k}{2} \log\bigl(\sigma^2\bigr) -
\frac{1}{2 \sigma^2} \sum_{i=1}^k
u_i^2
\end{eqnarray*}
denotes the ``complete-data log-likelihood,'' what the log-likelihood
would be if the random effects were observable. We consider
implementation of MCEM in a benchmark data set given by Booth and
Hobert [(\citeyear{boothobe1999}), Table 2], assuming the true
parameter value $\theta= (\beta, \sigma^2) = (5.0, 0.5)$. In this data
set $x_{ij} = j/15$ for each $j = 1,\ldots, m_i \equiv15$, for each $i
= 1,\ldots, k = 10$. Let $\tilde{\theta} = (4.0, 1.5)$. We can take as
an MCMC approximation of the $Q$-function the sample average of the
chain $ \{ l_c(\theta; y, u^{(t)})  \}$, that is,
\[
Q(\theta; \tilde{\theta}) \approx{\overline{l}_{C}}_{n}(
\theta; \tilde{\theta}):= \frac{1}{n} \sum_{t=1}^{n}
l_c\bigl(\theta; y, u^{(t)}\bigr),
\]
where $ \{ u^{(t)}\dvtx  t = 1, 2,\ldots,n  \}$ is a realization
of one of our five Markov chains with stationary density $h(u | y;
\tilde{\theta})$ as defined by (\ref{eqbenchmarktarget}). For the
sake of simplicity we will consider estimating the point
$Q(\tilde{\theta}; \tilde{\theta})$ rather than the entire function.
The mixing conditions on the Markov chain ensure the existence of a
CLT for the Monte Carlo error ${\overline{l}_{C}}_{n}(\tilde{\theta};
\tilde{\theta}) - Q(\tilde{\theta}; \tilde{\theta}) $ with the
variance of the asymptotic normal distribution denoted
$\sigma^{2}_{Q}$, which can be consistently estimated with the batch
means estimator $\hat{\sigma}^{2}_{Q}$
(\cite{joneharacaffneat2006}).\vspace*{1pt}

\begin{figure*}

\includegraphics{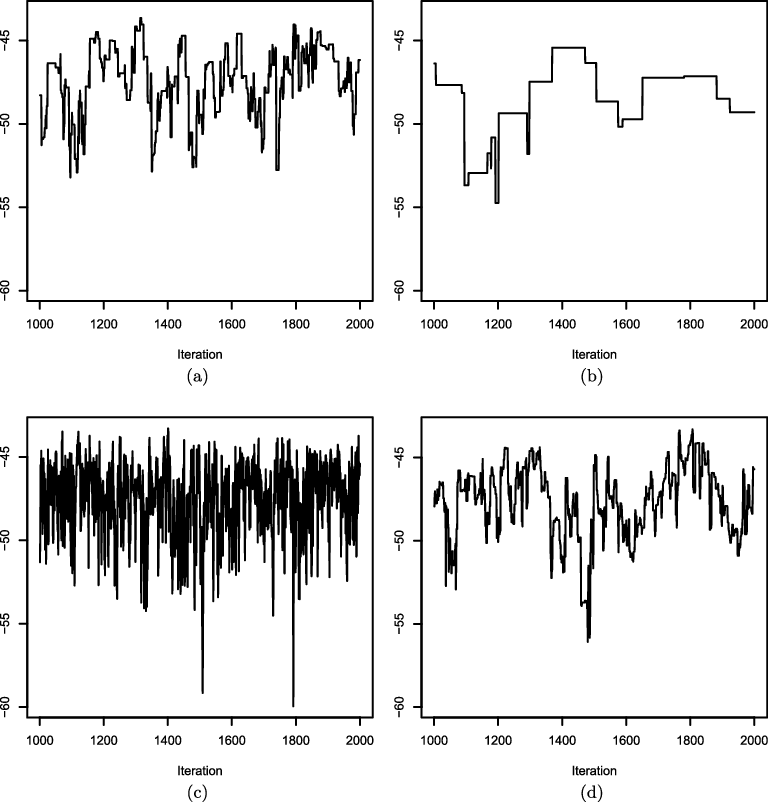}

\caption{Partial trace plots for the Markov chain $\{l_c(\theta; y,
u^{(t)} ) \}$ generated by \textup{(a)} RW, \textup{(b)} MHIS,
\textup{(c)} CIS and \textup{(d)} RSIS in the logit-normal example of
Section \protect\ref{secbenchmark}.} \label{figbenchmark}
\end{figure*}

We implemented MHIS, RW, CIS and RSIS as discussed above---we skip
reporting our implementation of RQIS, as it is very similar to CIS in
this example---in each case simulating a chain of length $n=10^6$ and
taking as our initial distribution  $U^{(0)} \sim\mathrm{N}_{10} (0,
\sigma^2 I)$. For the Metropolis random walk we drew our jump
proposals from a $\mathrm{N}_{10} (0, \tau^2 I)$, with $\tau^2 =
\sigma^2/6$ (this setting determined by trial and error, in order to
minimize the autocorrelation in the resulting chain, and yielded an
observed acceptance rate of $27.3\%$). A partial trace plot (the
second 1000 updates) is shown in panel (a) of
Figure~\ref{figbenchmark}. Analogous plots for the MHIS and
component-wise algorithms appear in the remaining panels.

Consider the trace plots for the four chains. The most striking
result is the dreadful performance of the MHIS, shown in panel (b).
The RW chain\break [panel~(a)] mixes much faster than the MHIS, but still
shows significant autocorrelation. Now RSIS\break [panel (d)] appears to
mix faster than MHIS but is very similar to RW. Finally, the CIS
[panel (c)] chain appears to be the best of these four samplers. This
suggests that when there is weak dependence between the components of
the target distribution, one should use a CWIS instead of a MHIS;
recall that \citet{nealrobe2006} reached a similar conclusion for
the Metropolis random walk.

In general, the empirical performance of MHIS depends entirely on the
``closeness'' of the proposal distribution to the target and, clearly,
the marginal distribution of the random effects $U$ is not
sufficiently similar to the conditional distribution of $U$ given the
data. It is worth recalling that, by
Theorem~\ref{thmdiscretemodels}, the MHIS depicted in panel (b) of
Figure~\ref{figbenchmark} is a uniformly ergodic Markov chain. Thus,
this example nicely illustrates the perils of over-reliance on
asymptotic properties of a sampler, which provide no guarantee of
favorable performance in finite-sample implementations.

\begin{table*}[t]
\caption{Results for the logit-normal example of
Section \protect\ref{secbenchmark}. Estimates of $Q(\theta; \tilde
{\theta})$,
and $\sigma^2_Q$ at $\theta= \tilde{\theta} = (4.0, 1.5)$, are
based on $n=10^6$; middle columns show half-width of 95\% confidence
interval ($t_* = 1.960$) and integrated autocorrelation time (ACT).
Rightmost panel shows estimated ESEJD based on $m=10^{3}$
replications, with standard errors in parentheses}
\label{tabex2}
\begin{tabular*}{\tablewidth}{@{\extracolsep{\fill}}ld{2.2}d{4.2}cd{3.2}c@{}}
\hline
\textbf{Algorithm} & \multicolumn{1}{c}{$\bolds{\hat{Q}(\theta| \tilde{\theta}; y)}$}
& \multicolumn{1}{c}{$\bolds{\hat{\sigma}^2_Q}$}
& \multicolumn{1}{c}{$\bolds{t_*\hat{\sigma}_Q/\sqrt{n}}$} & \multicolumn{1}{c}{\textbf{ACT}}
& \multicolumn{1}{c@{}}{$\bolds{\widehat{\mathrm{ESEJD}}}$}
\\
\hline
RW & -47.74 & 203.71 & 0.028 & 39.37 & 0.57 (0.0004) \\
MHIS & -47.77 & 2211.16 & 0.092 & 427.13 & 0.13 (0.0015) \\
CIS & -47.76 & 19.81 & 0.009 & 3.87 & 4.97 (0.0016) \\
RSIS & -47.76 & 258.85 & 0.032 & 50.08 & 0.50 (0.0005) \\
\hline
\end{tabular*}
\end{table*}

Consider the simulation results given in Table~\ref{tabex2}. Using
equivalent Monte Carlo sample sizes, the half-width of the interval
estimator is roughly the same for RW and RSIS. The half-widths for RW
and RSIS are more than 3 times larger than the half-width for CIS. On
the other hand, the half-width for MHIS is more than 3 times larger
than those of RSIS and RW and more than 10 times larger than that of
CIS. The ACTs tell a similar story, RW and RSIS are comparable while
MHIS is the worst and CIS is much better.

Estimation of ESEJD is based on the same $m=10^{3}$ independent
replications of RW, MHIS, RSIS and CIS for $n=10^{4}$ iterations
each. The results here are consistent with the other measures in the
above discussion. The performance of MHIS is terrible, while RSIS and
RW are comparable and CIS is the best of the four by a wide margin. The
fact that RSIS is comparable to RW is surprising. In RSIS only one of
the 10 components has a chance to be updated at each step, yet its
performance is similar to a chain which updates all of its components
about 30\% of the time.

\section{Concluding Remarks}
\label{secconclude}

Outside of the two-variable Gibbs sampler and the random scan
Metropolis-within-Gibbs algorithms,\break there has been little research on
convergence rates of component-wise MCMC samplers. This is
unfortunate because, as outlined in Section~\ref{secIntro},
establishing geometric ergodicity is a key step in enabling a
practitioner to have as much confidence in the simulation results as
if the samples were independent and identically distributed.

Certainly a theme of this paper has been that studying the convergence
rates of component-wise samplers formed by composition,\vadjust{\goodbreak} that is,
$P_{C}$, enables us to establish uniform or geometric ergodicity for
other component-wise samplers such as $P_{\RQ}$ and $P_{\mathrm{RS}}$. Indeed,
we showed this is true for uniform ergodicity in the general setting
and for geometric ergodicity in the two-variable setting. It seems
that studying the convergence rates of $P_{\RQ}$ and $P_{\mathrm{RS}}$ should
also inform us about the rate of $P_{C}$. Specifically, it is
tempting to think that $P_{\mathrm{RS}}$ should converge no faster than
$P_{\RQ}$ which should converge no faster than $P_{C}$. Especially in
the two-variable setting, we suspect this is the case. Indeed,
\citet{tanjonehobe2012} study a class of target distributions and
show that either both GS and RSGS are geometrically ergodic or neither
are.

Another theme has been that component-wise\break MCMC methods can be
superior to full-dimensional updates. For example,
full-dimensional
MCMC methods often fail to be geometrically ergodic, but obvious
component-wise implementations are. Also, the empirical investigations
in Section~\ref{secExamples} showed that the finite sample properties
of component-wise methods were superior to full-dimensional methods in
every case, which matches our observation in so many real data
examples; see the references in Section~\ref{secIntro}. The near
ubiquity of component-wise methods in the applied literature suggests
that this view is widely held among MCMC practitioners.

\section*{Acknowledgments}

Galin L. Jones supported in part by the National Science Foundation and
the National Institutes for Health. Ronald C. Neath was supported by a
PSC-CUNY Award, jointly funded by The Professional Staff Congress and
The City University of New York.

\begin{supplement}
\stitle{Supplementary material for ``Component-Wise\break Markov Chain Monte
Carlo: Uniform and Geometric Ergodicity Under Mixing and Composition''\\}
\slink[doi]{10.1214/13-STS423SUPP} 
\sdatatype{.pdf}
\sfilename{sts423\_supp.pdf}
\sdescription{This supplemental article includes all technical details
and proofs for the above results.}
\end{supplement}


\end{document}